\documentclass[preprint,secnumarabic,amssymb, aps, prl]{revtex4}

\usepackage{graphicx}
\usepackage{amsfonts,amsmath,amssymb}

\newcommand{\atan}{\arctan}
\begin{document}
\title{Current induced domain wall dynamics in the presence of a transverse magnetic field in out-of-plane magnetized materials}
\author{O. Boulle, L. D. Buda-Prejbeanu,  M. Miron and G. Gaudin} 
\affiliation{SPINTEC, CEA/CNRS/UJF/INPG, INAC, 38054 Grenoble Cedex 9, France}

\date{\today}%
\begin{abstract}
 An analytical model was developped to describe the current induced DW dynamics of a Bloch DW in the presence of an external transverse magnetic field.  The model  takes into account the DW deformation and the magnetization tilting in the domain.   The model is compared to the results of micromagnetic simulation  and an excellent agreement is obtained.  In the steady state regime, the model shows that the domain tilting does not change the DW mobility. An external or  current induced transverse magnetic field  such as the Oersted or Rashba  field can prevent the Walker breakdown leading to a higher domain wall velocity.

\end{abstract}

\maketitle
\section{Introduction}

The possibility to manipulate a domain wall (DW) using a spin polarized current    has opened a new path to write the  information in magnetic memory devices~\cite{Boulle11MSaERR}. Several innovative memory devices based on this effect have been recently proposed. One can cite    three-terminal MRAM schemes  where the different writing and reading path potentially solves the reliability issue observed in standard STT MRAM~\cite{Fukami08ITM,Cros05BI}. The magnetic racetrack where the information is stored in tens of DWs simultaneously moved by current pulses is also a potential disruptive technology for the mass storage market which offers a fast, cheap and solid state alternative to  hard disk drives~\cite{Parkin08S}. The key factors for such memories are  the write(/read) time and the power consumption of the device. An important effort of research has been carried out these last years to find out the ideal material with fast DW motion ($>100$ m/s) and low critical current density ($\sim10^{11}$~A/m$^2$). While most studies initially focused on the soft materials such as permalloy, the attention has now shifted to out-of-plane magnetized materials with strong perpendicular anisotropy where spin transfer was shown to be more efficient. These materials also have the advantage of narrow DWs ($\sim5$~nm) meaning a potentially very high  areal density. Besides the material issues, the nature of the DW dynamics plays also a major role.
Indeed, two dynamical regime are expected depending on the current density~\cite{Thiaville05EL} : at low current density, a steady state regime   with a stable DW structure  and at high current density, a turbulent regime   where the DW structure continuously oscillates. The limit between these two regimes is  the so-called Walker breakdown similar to the one observed when applying an easy-axis magnetic field. In large perpendicular anisotropy materials where high non-adiabatic effects have been identified~\cite{Boulle11MSaERR}, one expects a much higher DW mobility  in the steady state regime compared to the precessional regime.
However,  a   current density higher than the Walker current is generally needed for fast   DW motion  so the low mobility turbulent regime is reached. 
Recently, several schemes~\cite{Pizzini09APE,Uhlir11PRB,Miron10NM,Jang10JAP} have been proposed  to prevent the occurrence of this low mobility regime using in-plane magnetic field transverse to the wire where the DW propagates: the transverse field stabilizes the DW structure and thus extends the high mobility regime toward higher current density. This transverse magnetic field can be externally applied~\cite{Jang10JAP} or more advantageously induced by the current injection using the current Oersted field in multilayer geometry~\cite{Pizzini09APE,Uhlir11PRB,Jang12APL} or the Rashba field in inversion asymetry multilayer structure~\cite{Miron11NMa}. Whereas the effect of a transverse magnetic field on the field driven DW dynamics has been studied both experimentally~\cite{Bryan10IToM,Glathe10PRB,Jang10JAP,Lu10JAPa,Richter10APL,Seo10APLa,Bryan08JAP,Glathe08APLa,Kunz08JAP,You08APL,You08APLa} and theoretically~\cite{Hubert74,Sobolev95JMMM,Teh95JMMM,Kaczer61CJoP}, the effect on the current driven dynamics has been little addressed. Of particular interest is the case of out-of-plane magnetized materials where the DW dynamics can generally be well described by a  1D analytical  model due to the simple Bloch DW structure. 

Here we present an analytical model to describe the current induced DW dynamics of a Bloch DW in the presence of an external transverse magnetic field.  The model is based on a modified Bloch magnetization profile that takes into account the DW deformation and the magnetization tilting in the domain~\cite{Kaczer61CJoP,Hubert74}. 
The model gives a simple analytical expression of the current induced DW velocity in the presence of a transverse magnetic field  in the steady state regime and an excellent agreement with micromagnetic simulations is obtained. The model shows that the transverse magnetic field leads to an increase of the Walker current allowing high mobility DW motion even at high current densities.

The one-dimensional model  initially proposed by Slonczewski~\cite{Slonczewski79} and later extended to include the effect of spin torque~\cite{Thiaville02JMMM,Thiaville05EL}, provides a convenient and intuitive way to describe the DW dynamics in the presence of current. This model is based on the assumption that the DW keeps  its static structure during its motion  determined by the exchange and the anisotropy whereas the forces inducing the dynamics are first order correction affecting only the DW position $q$, the DW angle $\psi$ and the DW width $\Delta$. This model was shown to provide a good qualitative description of the DW dynamics of the N\'eel wall found in in-plane magnetized wires and  a  nearly quantitative description for Bloch wall~\cite{Szambolics09JMMM}. 

The dynamical 1D equations are based on the Landau-Lifschitz equation where the current induced spin torques have been added :
\begin{equation}\label{LLG}
\frac{\partial \textbf{m}}{\partial t}= -\gamma\frac{\delta E}{\delta\textbf{M}}\times \textbf{m} + \alpha \textbf{m} \times \frac{\partial \textbf{m}}{\partial t} - u\frac{\partial \textbf{m}}{\partial x}
+ \beta u\ \textbf{m} \times  \frac{\partial \textbf{m}}{\partial x}
\end{equation}

Here $\gamma=g|\mu_{B}|/\hbar$ is the gyromagnetic ratio and $E$ the free energy density. The third term is the adiabatic spin-transfer torque where $u=JPg\mu_B/2eM_s$,  $\mu_{B}$   the Bohr magneton,  $M_s$  the saturation magnetization, $J$ the current density. The fourth term is the non-adiabatic torque~\cite{Thiaville05EL} described by the parameter $\beta$.

In the following, two cases are distinguished depending on the amplitude of the external transverse magnetic field $H_t$ compared to the anisotropy field $H_{an}=2K_{an}/\mu_0M_s$.

\subsection{Low transverse magnetic field }

 For $H_t\ll H_{an}$, the DW structure and the magnetization in the domain are little affected by the transverse field and one can assume a standard Bloch profile. The polar and azimuthal angles  $\theta$ and $\varphi$ are then assumed as  $\theta=2\atan(\exp\{[x-q(t)]/\Delta(\psi)\})$ and   $\varphi=\psi(t)$ with $\psi(t)$ constant.    Here $\Delta(\psi)= \sqrt{A/\kappa}$ 
is the DW width with $\kappa=(K_{an}+K\sin^2 \psi)^{1/2}$ where $K_{an}=K_{an}^0-\mu_0M_s^2$ with $K_{an}^0$  the easy axis magnetic anisotropy,  $A$ the exchange constant,  $K=\mu_0M_s H_k/2$ with $H_k$ the DW demagnetizing field.  The integration of Eq.~\ref{LLG} over this DW profile leads to  the following 1D equations:

\begin{eqnarray}\label{EQ:1Dmodel}
\frac{\alpha}{\Delta}(\dot{q}-\frac{\beta }{\alpha}u)+\dot{\psi} & = &  -\frac{\gamma}{2M_s}\frac{\partial \sigma}{\partial q}\\
\dot{q}-u-\alpha\Delta\dot{\psi} & =  & \frac{\gamma}{2M_s}\frac{\partial\sigma}{\partial\psi}
\end{eqnarray}

 For $H_t\ll H_{an}$, the  DW energy per unit surface can be written as 
\begin{equation}
	\sigma=4\sqrt{A\kappa}+ H_kM_s\Delta\sin^2\psi-2 HM_sq-\pi\Delta M_sH_t\cos(\psi-\psi_t)
	 \label{sigma}
\end{equation}
with $H$ the easy external magnetic field and $\psi_t$ the in-plane direction of the transverse magnetic field\,\footnote{$\psi_t=0$ for the external magnetic field applied along the y axis}.
Eq.~\ref{sigma} then leads to the following 1D equations : 


\begin{eqnarray}
\label{1Dmodel1}
\frac{\alpha}{\Delta}(\dot{q}-\frac{\beta }{\alpha}u)+\dot{\psi} & = & \gamma H\\
\label{1Dmodel2}
\dot{q}-u-\alpha\Delta\dot{\psi} & = &  \frac{\gamma\Delta H_k }{2}\sin 2\psi+\frac{\pi\gamma\Delta H_t}{2}\sin(\psi-\psi_t)
\end{eqnarray}
The transverse magnetic field thus enters the equation as an additional torque on the DW proportional to $\sin(\psi-\psi_t)$.

\subsection{Large transverse magnetic field }
\begin{figure}[!h]
	\centering
		\includegraphics[width=1\textwidth]{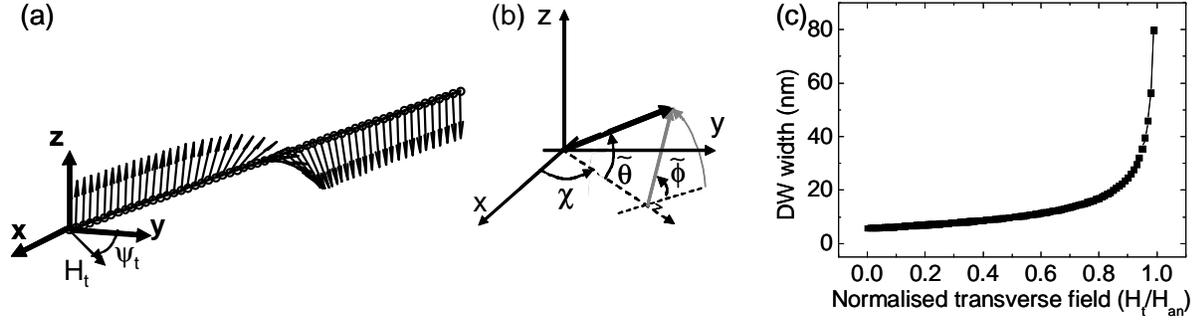}
	\caption{(a) Schematics of the DW profile. (b) Coordinate system related to $\tilde{\theta}$,  $\tilde{\varphi}$ and $\chi$. (c) DW width as a function of the transverse field amplitude applied along the y direction ($\psi_t=0$). The following magnetic and transport parameters are used corresponding to a Pt (3 nm)/Co (0.6 nm)/AlOx mutltilayer~\cite{Miron10NM} : $M_s=1.09\times10^6$ A/m,$A=1.6\times10^{-11}$ J/m$^2$, $K_{an}^0=1.25\times10^6$~J/m$^3$ ,  $H_k=15$ mT.}
	\label{Fig1}
\end{figure}

For larger $H_t$, the magnetization tilting in the domain as well as the DW structure deformation have to be taken into account. 
This problem was already tackled in the case of field driven DW dynamics by Hubert~\cite{Hubert74,Hubert98}. To account for the tilting of the magnetization in the domain and DW deformation, more convenient polar and azimuthal angles $\tilde{\theta}$ and $\tilde{\varphi}$ are defined (see Fig.~\ref{Fig1}). Their correspondent axes are rotated by an angle $\chi$ that defines the tilt of the  DW plane due to  DW dynamics. The application of the transverse magnetic field along a  direction $\psi_t$ leads to a tilt of the magnetization in the domain along $\tilde{\theta}=\tilde{\theta_0}$ and $\tilde{\varphi}=\pm\tilde{\varphi_0}$ with 
\begin{eqnarray}
	\cos\tilde\theta_0=h\sin(\psi_t- \chi)\\
	\sin\tilde\theta_0\cos\tilde\varphi_0=h\cos(\psi_t- \chi)
\end{eqnarray}
where $h=H_t/H_{an}$.
In between, the angle $\tilde{\varphi}$   rotates  as~\cite{Kaczer61CJoP,Hubert74} :
\begin{eqnarray}
\label{DWprofile}
\cos\tilde{\varphi}- \cos\tilde{\varphi_0}=\frac{\sin^2\tilde{\varphi_0}}{\cos\tilde{\varphi_0}+\cosh[k_0( \chi)(x-q(t))]}\\ k_0(\chi)=\sin\tilde{\varphi_0}\sqrt{(K_{an}+K\sin^2 \chi)/A}\\
\end{eqnarray}

The application of the transverse magnetic field leads to a widening of the DW (see Fig.~\ref{Fig1}(c)) with a DW width scaling with $h$  as~\cite{Hubert74}  :

\begin{equation}
	\Delta_t(h)=\frac{2\sqrt{A/K_{an}}\sqrt{1-h^2}}{[\sqrt{1-h^2\sin^2(\psi_t-\chi)}-h\cos(\psi_t-\chi)]\sqrt{1+K_d/K_{an}\sin^2\chi}}
\end{equation}
\\
To describe the current induced DW dynamics for such a DW profile, a Lagrangian approach is considered. Indeed, the LLG equation can be derived by applying the Lagrange-Rayleigh equation~\cite{Hubert74,Thiaville02JMMM,Thiaville05EL,Lucassen09PRB} 
 \begin{equation}
 \frac{\partial L}{\partial X}-\frac{\partial  }{\partial  t} \frac{\partial L}{\partial \dot{X}}-\frac{\partial }{\partial x}\frac{\partial L}{\partial X'}  +\frac{\partial F}{\partial \dot{X}}=0
 \label{Lagrange}
 \end{equation}

 with X'=$\partial X/\partial x$  and using  the Lagrangian $Lt$ :

\begin{equation}
L_t=E+\frac{M_s}{\gamma}(\cos\bar{\theta}-\cos\bar{\theta_0})-(uM_s/\gamma)\bar{\varphi}\mathrm{d}(\cos\bar{\theta})/\mathrm{dx}
\end{equation}
where $E$ the micromagnetic energy density and $\bar{\theta}$ and  $\bar{\varphi}$  the polar and azimuthal angles  defined as $\textbf{m}=(\cos\bar{\theta},\sin\bar{\theta}\cos\bar{\varphi},\sin\bar{\theta}\sin\bar{\varphi}$). The angles $\bar{\theta_0}$ ans $\bar{\varphi_0}$ describes the magnetization tilting in the domain with $\cos\bar\theta_0=h\sin\psi_t$ and $\sin\bar\theta_0\cos\bar\varphi_0=h\cos\psi_t$.   The Lagrangian is chosen so that it cancels out in the domain. The effect of the damping and the non-adiabatic torque is included in  the  dissipative function $F$~\cite{Lucassen09PRB} :  
\begin{equation}
\label{Eqdiss}
F=\frac{\alpha M_s}{2\gamma}\left[(\frac{d}{dt}+\frac{\beta u}{\alpha}\frac{d}{dx})\textbf{m}\right]^2
\end{equation}
The integration of $L_t$ and the dissipation function $F$   over the DW profile  of Eq.~\ref{DWprofile} leads to~\cite{Hubert74} :

\begin{eqnarray}
	L_{DWt} & = & \sigma-(2M_s/\gamma)(\dot{q}-u)f_k\\
	F_{DWt}& = & \frac{\alpha M_s}{\gamma}\frac{E_G}{4A}\left[(\dot{q}-\frac{\beta}{\alpha}u)^2+\frac{\dot{\chi}^2}{k_0^2}+\dot{\chi}^2\left(\frac{\partial 1/k_0}{\partial \chi}\right)^2\frac{1}{3}(\pi^2-\tilde{\varphi_0}^2-\frac{2\tilde{\varphi_0}^2}{1-\tilde{\varphi_0}\mathrm{cotan}\tilde{\varphi_0}})\right]
\end{eqnarray}
where 
\begin{equation}
f_k=\frac{1}{2}\int_{-\varphi_0}^{\bar{\varphi_0}}(\cos\bar\theta-\cos\bar\theta_0) d\bar\varphi
\end{equation}

\begin{multline}
E_G=4\sqrt{AK_{an}}\sqrt{(1-h^2\sin^2(\psi_t-\chi))(1+K_0/K_{an}\sin^2\chi)}\\
\left[\sqrt{1-h^2}-h\cos(\psi_t-\chi)\arctan\frac{\sqrt{1-h^2}}{h\cos(\psi_t-\chi)}\right]
\end{multline}

Note that  the adiabatic torque leads to the  operation $\dot{q}\rightarrow\dot{q}-u$ in the Lagrangian whereas the non-adiabatic torque leads to the operation $\dot{q}\rightarrow\dot{q}-\beta u/\alpha $ in the dissipation function.
The Lagrange-Rayleigh equation with $X=\chi$ or $q$ leads to the following 1D equations :

\begin{eqnarray}
\label{Eq:vsteady1}
\alpha\frac{E_G}{4A}(\dot{q}-\frac{\beta}{\alpha}u)+\frac{\partial f_k}{\partial\chi}\dot{\chi}  =   \gamma H \sqrt{1-h^2}\\
\label{Eq:Vsteady2}
\frac{\partial f_k}{\partial\chi}(\dot{q}-u)-\frac{\alpha E_G}{4A}\left[\frac{1}{k_0^2}+\left(\frac{\partial 1/k_0}{\partial \chi}\right)^2\frac{1}{3}(\pi^2-\tilde{\varphi_0}^2-\frac{2\tilde{\varphi_0}^2}{1-\tilde{\varphi_0}\mathrm{cotan}\tilde{\varphi_0}})\right]\dot{\chi} =  \frac{\gamma}{2M_s}\frac{\partial E_G}{\partial \chi}
\end{eqnarray}

From Eq.~\ref{Eq:vsteady1}, one can easily derive the   DW velocity in the steady state ($\dot{\chi}=0$):
\begin{equation}
\label{Vsteady2}
	\dot q=\beta u/\alpha + \frac{\gamma 4A}{\alpha E_G} H \sqrt{1-h^2}
\end{equation}

to be compare to the steady state DW velocity in the absence of transverse magnetic field obtained from Eq.~\ref{1Dmodel1} :

\begin{equation}
	\dot q=\beta u/\alpha+\frac{\gamma \Delta}{\alpha} H  
\end{equation}

An important result is thus that the  DW mobility induced by the current ($\beta u /\alpha$) is not affected by the transverse magnetic field in the steady state regime. On the contrary, the DW mobility induced by the easy axis field $H$ is increased when applying the transverse magnetic field  due to the large decrease in the DW energy $E_G$ when $H_t$ increases.

\begin{figure}[!h]
	\centering
		\includegraphics[width=1.00\textwidth]{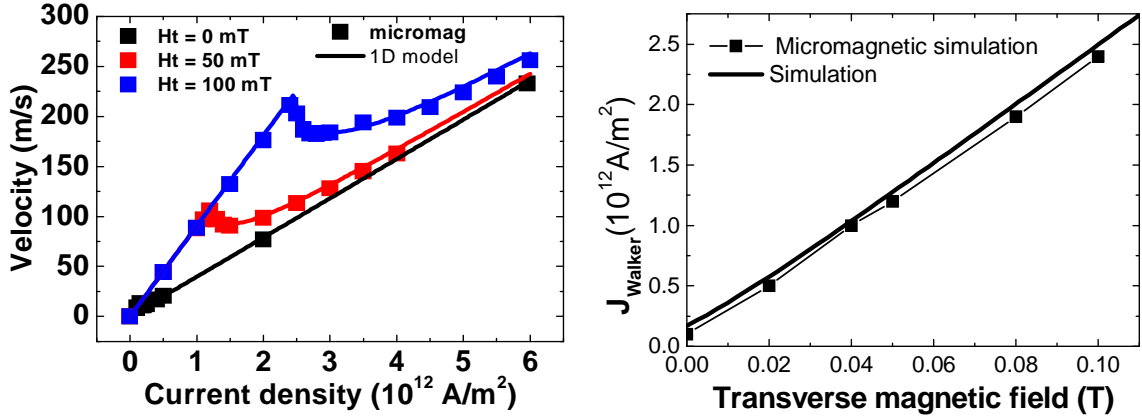}
	\caption{(a) DW velocity as a function of the current density for different transverse magnetic field applied ($\psi_t=0$) predicted by the 1D model (plain lines) and micromagnetic simulation (square plots). (b) Walker current as a function of the transverse magnetic field predicted by the 1D model (square) and micromagnetic simulation (line). The following magnetic and transport parameters are used: $M_s=1.09\times10^6$ A/m,$A=1.6\times10^{-11}$ J/m$^2$, $K_{an}^0=1.25\times10^6$~J/m$^3$ ,  $H_k=15$ mT, $\alpha=0.5$, $\beta=1.7$. }
	\label{Fig2}
\end{figure}

To obtain the DW velocity above the Walker breakdown, numerical resolutions of Eq.~\ref{Eq:vsteady1} and Eq.~\ref{Eq:Vsteady2} have been carried out   (Fig.~\ref{Fig2}(a), plain line). The main effect of the application of the transverse magnetic field ($\psi_t=0$) is an increase of the Walker current: the transverse field reduces the DW tilt angle and thus  shifts the Walker current toward high current.   Below the Walker current, the velocity follows $v=\beta u/\alpha$ as expected, whereas in the precession regime well above the Walker breakdown, the velocity gets close to the zero transverse field velocity $<v>=(1+\alpha\beta)/(1+\alpha^2)u$. The increase of the Walker current with the transverse magnetic field is shown on Fig~\ref{Fig2}(b). 

\begin{figure}[!h]
	\centering
		\includegraphics[width=0.8\textwidth]{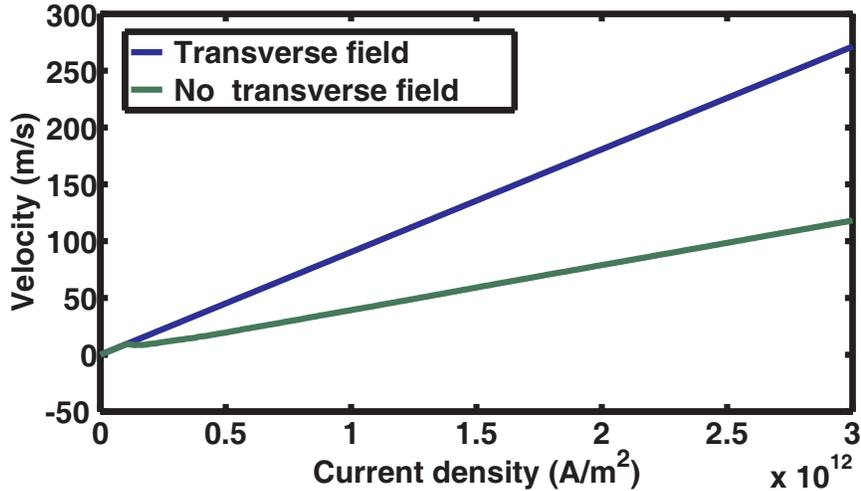}
	\caption{(a) DW velocity as a function of the injected current density at zero external transverse field and in the presence of a current induced transverse field (0.1~T for $10^{12}$~A/m$^2$) along $\psi_t$=0. The other parameters are the same as Fig.~\ref{Fig2}.}
	\label{Fig3}
\end{figure}

To test the validity of the model, micromagnetic simulations have been carried out using the micromagnetic solver DWmicro3D based on the finite differences approximation for a 100 nm  wide nanowire and 0.6 nm thick Co layer~\cite{Szambolics09JMMM}. A high perpendicular anisotropy is assumed  leading to an out-of-plane magnetization and narrow Bloch DWs. The  calculated velocity as a function of the current density and the Walker current are plotted   in Fig.~\ref{Fig2} (square symbols).  One can see that an excellent agreement is obtained between the 1D model and the micromagnetic simulations. Such a good agreement is generally not observed for in-plane magnetized N\'eel DW and  can be accounted for to the rigid Bloch DW structure in this narrow wire. The profile of the magnetization transverse to the wire shows  little dependence of the magnetization on the direction transverse to the wire. The DW thus behave effectively as a rigid one-dimensional object.

The model can also be used to simulate the effect of a current induced transverse magnetic field, such as an Oersted field or a Rashba field. Fig.~\ref{Fig3} shows the velocity as a function of the current density for a transverse magnetic field of 0.1~T/10$^{12}$~A/m$^2$. As expected, the current induced transverse field stabilizes the DW structure and thus prevents the  Walker breakdown. The DW stays in the higher mobility steady state regime and a higher velocity is obtained.


In conclusion, a one-dimensional model of the current induced domain wall dynamics in the presence of a transverse magnetic field that takes into account the DW deformation and magnetization tilting in the domains was developed. In the steady state regime, the model shows that the domain tilting does not change the DW mobility.   The transverse magnetic field leads to an increase of the Walker current and thus the high mobility regime is extended toward  high current densities. A current induced transverse magnetic field such as Oersted or Rashba can prevent the Walker breakdown leading to a higher domain wall velocity. This work was  financially supported by the ANR 11-B510-008 ESPERADO project.

\appendix*
\section{Calculation of the kinetic energy}
$f_k$ is proportionnel to the kinetic energy of the DW $E_{kin}$~\cite{Hubert74} with :

\begin{equation}
E_{kin}=-\frac{2M_s\dot q }{\gamma}f_k
\end{equation}

 It can be expressed in the ($\tilde\theta,\tilde\varphi$) coordinate system as:
\begin{eqnarray}
f_k=\int_0^{\tilde\varphi_0}(\cos\bar\theta-h\sin\psi_t)(\cos\chi+g(\tilde\theta,\tilde\varphi)\sin\chi )d\tilde\varphi
\end{eqnarray}
where
\begin{eqnarray}
g(\tilde\theta,\tilde\varphi) & = & (\cos\tilde\varphi\sin\tilde\theta_0\cos\chi-\cos\tilde\theta_0\sin\chi)\cos\bar\theta/(1-\cos\bar\theta^2)\\
\cos\bar\theta & = & \cos\tilde\varphi\sin\tilde\theta_0\sin\chi+\cos\tilde\theta_0\cos\chi\\
\cos\tilde\theta_0 & = & h\sin(\psi-\chi)\\
\tan\tilde\varphi_0 & = & \frac{\sqrt{1-h^2}}{h\cos(\psi-\chi))}
\end{eqnarray}
For numerical calculation, an analytical expression of $f_k$ was obtained using the commercial software Mathematica.


\begin{thebibliography}{10}

\bibitem{Boulle11MSaERR}
Boulle, O., Malinowski, G., and Kl\"{a}ui, M.
\newblock {\em Materials Science and Engineering: R: Reports}{ \bf 72}(9),
  159--187 September  (2011).

\bibitem{Fukami08ITM}
Fukami, S., Suzuki, T., Ohshima, N., Nagahara, K., and Ishiwata, N.
\newblock {\em IEEE Trans. Mag.}{ \bf 44}(11), 2539 --2542 nov.  (2008).

\bibitem{Cros05BI}
Cros, V., Grollier, J., Sancher, M.~M., Fert, A., and Dau, F. N.~V.
\newblock {\em Brevet International}{ \bf }, WO 2006/064022 (2005).

\bibitem{Parkin08S}
Parkin, S. S.~P., Hayashi, M., and Thomas, L.
\newblock {\em Science}{ \bf 320}(5873), 190 -- 194 (2008).

\bibitem{Thiaville05EL}
Thiaville, A., Nakatani, Y., Miltat, J., and Suzuki, Y.
\newblock {\em Europhys. Lett.}{ \bf 69}, 990 (2005).

\bibitem{Pizzini09APE}
Pizzini, S., Uhlir, V., Vogel, J., Rougemaille, N., Laribi, S., Cros, V.,
  Jimenez, E., Camarero, J., Tieg, C., Bonet, E., et~al.
\newblock {\em Appl. Phys. Express}{ \bf 2}(2), 3003 (2009).

\bibitem{Uhlir11PRB}
Uhlí?, V., Pizzini, S., Rougemaille, N., Cros, V., Jiménez, E., Ranno, L.,
  Fruchart, O., Urbánek, M., Gaudin, G., Camarero, J., Tieg, C., Sirotti, F.,
  Wagner, E., and Vogel, J.
\newblock {\em Phys. Rev. B}{ \bf 83}(2), 020406 January  (2011).

\bibitem{Miron10NM}
Miron, I.~M., Gaudin, G., Auffret, S., Rodmacq, B., Schuhl, A., Pizzini, S.,
  Vogel, J., and Gambardella, P.
\newblock {\em Nat. Mater.}{ \bf 9}(3), 230--234 March  (2010).

\bibitem{Jang10JAP}
Jang, Y., Yoon, S., Lee, S., Lee, K., and Cho, B.~K.
\newblock {\em J. Appl. Phys.}{ \bf 108}(6), 063904 (2010).

\bibitem{Jang12APL}
Jang, Y., Mascaro, M.~D., Beach, G. S.~D., and Ross, C.~A.
\newblock {\em Appl. Phys. Lett.}{ \bf 100}(11), 112401--112401--5 March
  (2012).

\bibitem{Miron11NMa}
Miron, I.~M., Moore, T., Szambolics, H., {Buda-Prejbeanu}, L.~D., Auffret, S.,
  Rodmacq, B., Pizzini, S., Vogel, J., Bonfim, M., Schuhl, A., and Gaudin, G.
\newblock {\em Nat. Mater.}{ \bf 10}(6), 419--423 June  (2011).

\bibitem{Bryan10IToM}
Bryan, M.~T., Schrefl, T., and Allwood, D.~A.
\newblock {\em {IEEE} Transactions on Magnetics}{ \bf 46}(5), 1135--1138 May
  (2010).

\bibitem{Glathe10PRB}
Glathe, S., Zeisberger, M., Hübner, U., and Mattheis, R.
\newblock {\em Phys. Rev. B}{ \bf 81}(2) (2010).

\bibitem{Lu10JAPa}
Lu, J. and Wang, X.~R.
\newblock {\em J. Appl. Phys.}{ \bf 107}(8), 083915 (2010).

\bibitem{Richter10APL}
Richter, K., Varga, R., {Badini-Confalonieri}, G.~A., and Va?zquez, M.
\newblock {\em Appl. Phys. Lett.}{ \bf 96}(18), 182507 (2010).

\bibitem{Seo10APLa}
Seo, S., Lee, K., Jung, S., and Lee, H.
\newblock {\em Appl. Phys. Lett.}{ \bf 97}(3), 032507 (2010).

\bibitem{Bryan08JAP}
Bryan, M.~T., Schrefl, T., Atkinson, D., and Allwood, D.~A.
\newblock {\em J. Appl. Phys.}{ \bf 103}(7), 073906 (2008).

\bibitem{Glathe08APLa}
Glathe, S., Berkov, I., Mikolajick, T., and Mattheis, R.
\newblock {\em Appl. Phys. Lett.}{ \bf 93}(16), 162505 (2008).

\bibitem{Kunz08JAP}
Kunz, A. and Reiff, S.~C.
\newblock {\em J. Appl. Phys.}{ \bf 103}(7), 07D903--07D903--3 January  (2008).

\bibitem{You08APL}
You, C.
\newblock {\em Appl. Phys. Lett.}{ \bf 92}(19), 192514--192514--3 May  (2008).

\bibitem{You08APLa}
You, C.
\newblock {\em Appl. Phys. Lett.}{ \bf 92}(15), 152507--152507--3 April
  (2008).

\bibitem{Hubert74}
H\"{u}bert, A.
\newblock {\em Theorie der Dom\"{a}nenw\"{a}nde in Geordneten Medien.}
\newblock Springer,  (1974).

\bibitem{Sobolev95JMMM}
Sobolev, V.~L., Huang, H.~L., and Chen, S.~C.
\newblock {\em J. Magn. Magn. Mat.}{ \bf 147}(3), 284--298 (1995).

\bibitem{Teh95JMMM}
Teh, C.~T., Sobolev, V.~L., and Huang, H.~L.
\newblock {\em J. Magn. Magn. Mat.}{ \bf 145}(3), 382--384 March  (1995).

\bibitem{Kaczer61CJoP}
Kacz\'{e}r, J. and Gemperle, R.
\newblock {\em Czechoslovak Journal of Physics}{ \bf 11}, 157--170 March
  (1961).

\bibitem{Slonczewski79}
Slonczewski, J. and A.P, M.
\newblock {\em Magnetic domain walls in bubble materials}.
\newblock Academic Press,  (1979).

\bibitem{Thiaville02JMMM}
Thiaville, A., Garc, J.~M., and Miltat, J.
\newblock {\em J. Magn. Magn. Mater.}{ \bf 242}, 1061 -- 1063 (2002).

\bibitem{Szambolics09JMMM}
Szambolics, H., Toussaint, J.~C., Marty, A., Miron, I.~M., and Buda-Prejbeanu,
  L.~D.
\newblock {\em J. Magn. Magn. Mat.}{ \bf 321}(13), 1912 -- 1918 (2009).

\bibitem{Hubert98}
Hubert, A. and Sch\"afer, R.
\newblock {\em Magnetic domains}.
\newblock Spinger-Verlag,  (1998).

\bibitem{Lucassen09PRB}
Lucassen, M.~E., van Driel, H.~J., Smith, C.~M., and Duine, R.~A.
\newblock {\em Physical Review B}{ \bf 79}(22), 224411--11 (2009).

\end{thebibliography}
\end{document}